\documentclass{article}

\usepackage{arxiv}

\usepackage[utf8]{inputenc} 
\usepackage[T1]{fontenc}    
\usepackage{hyperref}       
\usepackage{url}            
\usepackage{booktabs}       
\usepackage{amsfonts}       
\usepackage{nicefrac}       
\usepackage{microtype}      
\usepackage{cleveref}       
\usepackage{lipsum}         
\usepackage{graphicx}
\usepackage{natbib}
\usepackage{doi}

\setcitestyle{numbers}

\title{The GA4GH Task Execution API: Enabling Easy Multi Cloud Task Execution}

\newif\ifuniqueAffiliation
\uniqueAffiliationtrue

\author{ {\hspace{1mm}Alexander Kanitz} \\
	University of Basel \\
	Spitalstrasse 41, 4056 Basel, Switzerland \\
	\And
	{\hspace{1mm}Matthew H. McLoughlin} \\
	Genomics Team, Microsoft Research \& AI \\
	Redmond, WA,  98052, USA \\
	\And
	{\hspace{1mm}Liam Beckman} \\
	Department of BioMedical Engineering\\
	Oregon Health and Science University\\
	Portland, OR 97229\\
	\And
	{\hspace{1mm}The GA4GH Cloud Workstream} \\
	Global Alliance for Genomics and Health (GA4GH) \\
	\And
	{\hspace{1mm}Venkat S. Malladi} \\
	Genomics Team, Microsoft Research \& AI \\
	Redmond, WA,  98052, USA \\
	\And
	{\hspace{1mm}Kyle P.~Ellrott} \\
	Department of BioMedical Engineering\\
	Oregon Health and Science University\\
	Portland, OR 97229\\
	\texttt{ellrott@ohsu.edu} \\
}

\renewcommand{\shorttitle}{The GA4GH TES API}

\hypersetup{
pdftitle={The GA4GH TES API},
pdfsubject={q-bio.NC, q-bio.QM},
pdfauthor={Kyle Ellrott},
pdfkeywords={API Standards, HPC/Cloud Hybrid Computing},
}

\begin{document}
\maketitle

\begin{abstract}
    The Global Alliance for Genomics and Health (GA4GH) Task Execution Service (TES) API is a 
    standardized schema and API for describing and executing batch execution tasks. It provides a 
    common way to submit and manage tasks to a variety of compute environments, including on premise 
    High Performance Compute and High Throughput Computing (HPC/HTC) systems, Cloud computing platforms, 
    and hybrid environments. The TES API is designed to be flexible and extensible, allowing it to be 
    adapted to a wide range of use cases, such as “bringing compute to the data” solutions for federated 
    and distributed data analysis or load balancing across multi cloud infrastructures. This API has been 
    adopted by a number of different service providers and utilized by several workflow engines. Using its 
    capabilities, genomes research institutes are building hybrid compute systems to study life science.  
\end{abstract}


\section{Introduction}
In the field of bioinformatics and computational biology, and specifically the area of genomic computing, analysis of 
data is done by chaining together sets of command line programs. There is no common programming language level API 
because each of these tools is the product of long research in different areas of specialization written by different 
teams working at different institutions.  These tools may be decades old or still in the development phase. Each of 
these tools may also have vastly different software stacks with different library dependencies. 
To deal with this situation, several groups have developed workflow engines that describe how these tasks 
are chained together. In conjunction to this, biological data is being generated by a number of different 
institutions and being stored at service providers including private institutional computing and commercial cloud providers. 
With the average Whole Genome Sequencing file being more than 200GB, downloading all of the data to a single storage site is not 
practical. The Global Alliance for Genomics and Health (GA4GH) Task Execution Service (TES) API is an API for describing and executing 
batch execution tasks, written to allow researchers to easily submit to and manage tasks on a variety of compute environments. GA4GH is 
an international community that seeks to advance human health by enabling analysis of genomic and health-related data. Part of this effort 
includes the development of standards and frameworks to promote the secure, responsible, and effective use of genomic data. The TES API is 
one of the cloud capable APIs developed by the GA4GH Cloud workstream. Other APIs developed in this workstream include the Tool Registry 
Service (TRS), the Workflow Execution Service (WES) and the Data Registry Service (DRS) API specifications.

The TES API was designed to address on-premises High Performance Computing and High Throughput Computing 
(HPC/HTC) systems, cloud computing platforms, and hybrid environments. The TES API is designed to be flexible 
and extensible, allowing it to be adapted to a wide range of use cases. It can be used to execute simple tasks, 
such as running a single command on a single compute node, or as a backend for workflow management systems that are 
able to schedule the execution of individual tasks out of complex workflows involving multiple tasks and dependencies.

There have been limited previous efforts to standardize batch system interfaces, such as  DRMAA (Distributed Resource 
Management Application API). This standard was developed in 2007 as an Open Grid Forum (OGF) API specification for 
distributed resource management (DRM) systems, such as a cluster or grid computing infrastructure. While supported 
by a number of HPC queuing systems, the specification was expressed as a C library, requiring software to have 
library-level bindings to access the API. This effort did not specify an interface for network-based clients. 
Globus Grid Resource Allocation and Management (GRAM) is a federated API for connecting to remote job schedulers. 
However, it is a single platform solution infrastructure, and not available to other systems outside of Globus.

Kubernetes (K8s) is an open-source container orchestration platform widely used for managing 
containerized applications. While Kubernetes is commonly employed in both cloud and on premise 
environments, its primary concern is container orchestration through microservices. While Kubernetes 
offers powerful orchestration capabilities, for batch computing these capabilities are limited. The 
Kubernetes Jobs API does project capacity for batch execution; however, it does not manage large file 
transfer and is limited to Kubernetes-specific environments.

TES, on the other hand, was defined as an OpenAPI defined specification for HTTP servers, 
based on the OpenAPI specification, allowing network access to the system, as well as lightweight 
library bindings to rapidly be generated for new programming languages. The flexibility of the 
TES API, and its ability to be deployed in a number of different infrastructures is a valuable 
tool for the genomics community and other audiences that benefit from a cross-platform, 
cross-cloud batch execution solution. TES can help to simplify and streamline the execution of 
computational workflows. It can also help to reduce the cost of running these workflows by making 
it possible to use a variety of compute resources, including cloud computing platforms (Figure 1). 
The ecosystem currently includes implementations that support HPC/HTC cluster-based job schedulers, 
such as SLURM and Grid Engine, Kubernetes-based systems and commercial cloud-specific APIs including 
AWS Batch and Azure Batch. Multiple workflow engines have begun to utilize the API to deploy workloads on 
TES-fronted systems, including Galaxy\cite{Jalili_Afgan_Taylor_Goecks_2020}, 
Cromwell\cite{Voss_Van_der_Auwera_Gentry_2017}, 
Nextflow\cite{Di_Tommaso_Chatzou_Floden_Barja_Palumbo_Notredame_2017} and 
Snakemake\cite{Mölder_Jablonski_Letcher_Hall_Tomkins-Tinch_Sochat_Forster_Lee_Twardziok_Kanitz_et_al._2021}.

\begin{figure}[t]
\includegraphics[width=16cm]{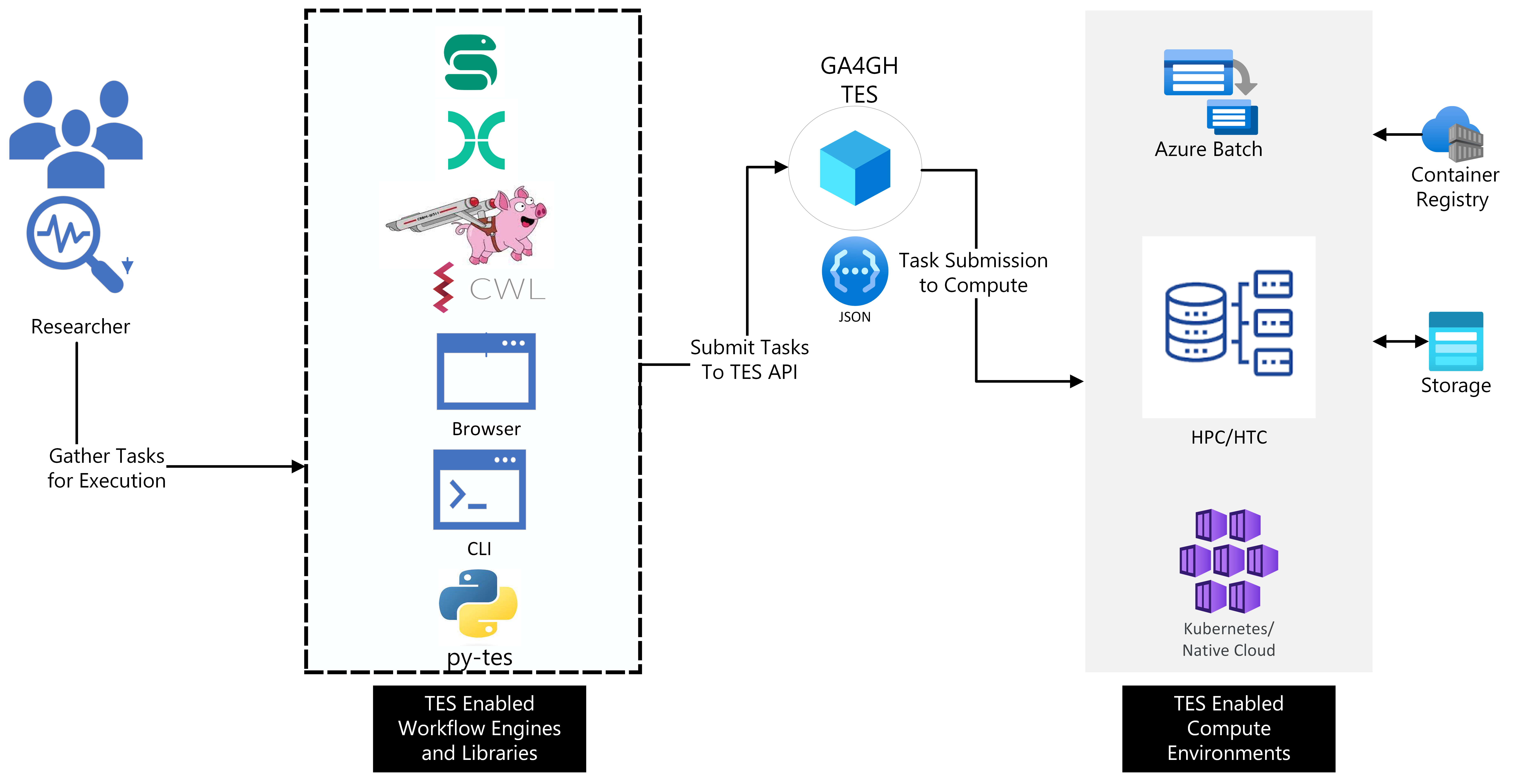}

Figure 1: Common TES use cases. The TES API wraps around compute environments providing a standard way of executing tasks.  Researchers can write and package their tasks and data in a domain-specific language (DSL) workflow language. They then hand the orchestration of the tasks over to the respective workflow management systems. The workflow management systems can then make use of TES clients to distribute tasks across different environments.  Alternatively, users can submit individual tasks to TES servers directly via command-line (CLI) or graphical user (GUI) interfaces. Thus, TES makes it easier for researchers to make use of a variety of compute environments seamlessly.Applications can support new compute environments by integrating with TES API, rather than develop unique connections for each environment. 
\end{figure}

\section{The TES Standard}
\label{sec:headings}

At the core of the TES API is the idea that in order for a new computational task to be issued into a compute 
environment, a number of core elements need to be defined so that systems can be adequately prepared for user 
code to be deployed. In a hybrid compute environment, there are a number of elements that must be explicitly 
defined to ensure that a command line is able to execute successfully. In many HPC/HTC environments, while a 
program command may be executed on an arbitrary node in the cluster, the system is usually set up in the exact 
same way as the submission host, including the same environmental variables, software libraries and support programs. 
HPC/HTC environments will also usually have global access to a unified POSIX file system, e.g., via the Network 
File System (NFS). In a cloud environment, no such guarantees exist. A process may be initiated on a virtual machine 
that was just allocated and initialized based on a base OS image with no additional software installed. Additionally, 
there is no guarantee that any particular file assets will be available, or that shared storage will be mounted. The 
core resource of the API is a Task (Figure 2), which is an atomic message that details all of the required task parameters. 
Because such guarantees are lacking in the execution environment, the TES specification requires a number of task parameters 
to be defined. These include: 

\begin{figure}[t]
\includegraphics[width=10cm]{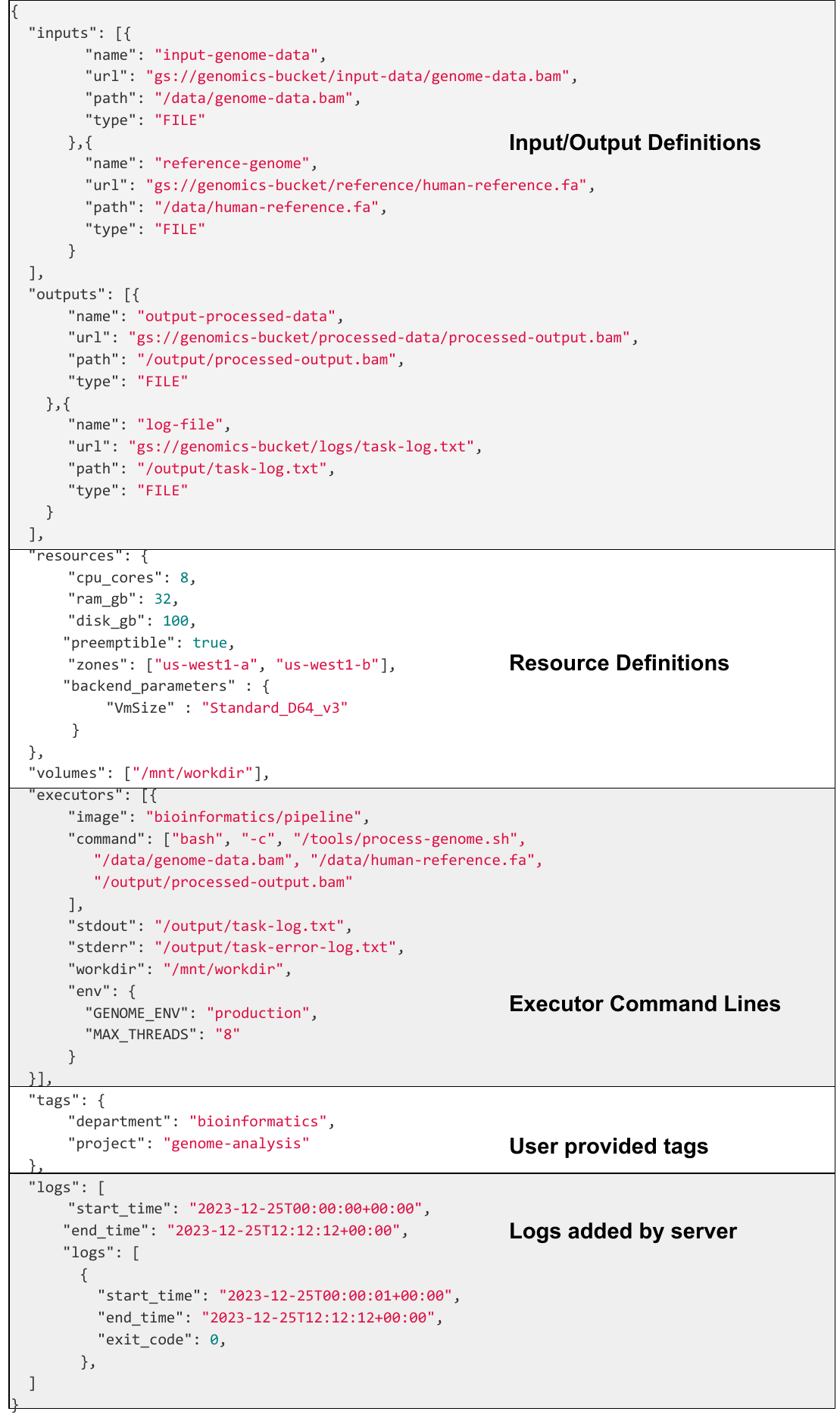}

Figure 2: Example TES packet.  An example TES task demonstrating the use of inputs, outputs, and logging
\end{figure}

\textbf{Application Environment}: User software and computational tools need to be packaged and deployed. As user jobs 
will be handled by potentially undefined systems, such as vanilla virtual machines, or HPC worker nodes with different 
operating system versions and hardware architectures, the software required to run the job needs to be defined as part of 
the job. Software containerization systems, such as Docker images, have provided a mechanism to capture entire software 
environments, with all software dependencies, and rapidly deploy them on new computer systems.

\textbf{Computational Resources}: Tasks have defined hardware minimums to be able to run. This includes CPUs, GPUs, memory 
and storage within a computing environment. It plays a crucial role in optimizing task execution by allocating resources 
dynamically based on task demands as provided in the API. The computational resources used for the job may be reclaimed and 
reset to be available for the next job. 

\textbf{Inputs}: Custom input files and directories for a specific task need to be transferred into place prior to the 
user task being invoked. These files need to be stored within a system that can be accessed by both the submission system 
and the eventual compute node. If all processes are maintained within the same HPC environment, this can be represented 
using traditional network attached storage, such as an NFS instance. In the case of cloud environments, Object Storage 
systems, such as S3, can be used as the intermediate storage instead.

\textbf{Outputs}: Next to requiring custom inputs,  most jobs will create one or more output files and directories that 
need to be exported to a common storage system at the end of the job. As any files not included in an export manifest 
may be lost after the end of a job.

\textbf{Environmental Variables}: A tool's behavior may be modified based on the settings of environmental variables, 
such as the current working directory. TES therefore supports the definition of such variables.

\textbf{Command Lines}: The user process will need a command line to be executed in the environment set up by the TES 
service as a result of the information stated above. TES supports the definition of multiple command lines for each 
request, e.g., in order to define setup and teardown jobs for the main job, or for the grouping of multiple command lines 
with similar requirements. When multiple command lines are defined in a TES request, they will be executed sequentially. 
Each command can utilize a different container image, but inputs, outputs and  volumes are shared across commands.

TES-compatible clients and servers can be developed and used independently. This makes it easy to integrate the TES API into 
existing workflow management systems and other software tools. Because the TES task messages are atomic, containing all 
information needed to successfully complete a set of command lines, they may be easily passed between clients, servers, 
proxy servers, databases and event streams. Given this definition of task, we have provided a frame for TES tasks that can be 
executed on a TES-compatible server, with the details of HPC/HTC vs cloud compute environment abstracted away (Figure 3). This 
makes it easy to port workflow processes to new compute resources. In practice, by providing this core API the TES API is 
flexible enough to be used to execute a wide range of tasks, from simple commands to interacting with engines to orchestrate 
complex workflows.

\begin{figure}[t]
\includegraphics[width=10cm]{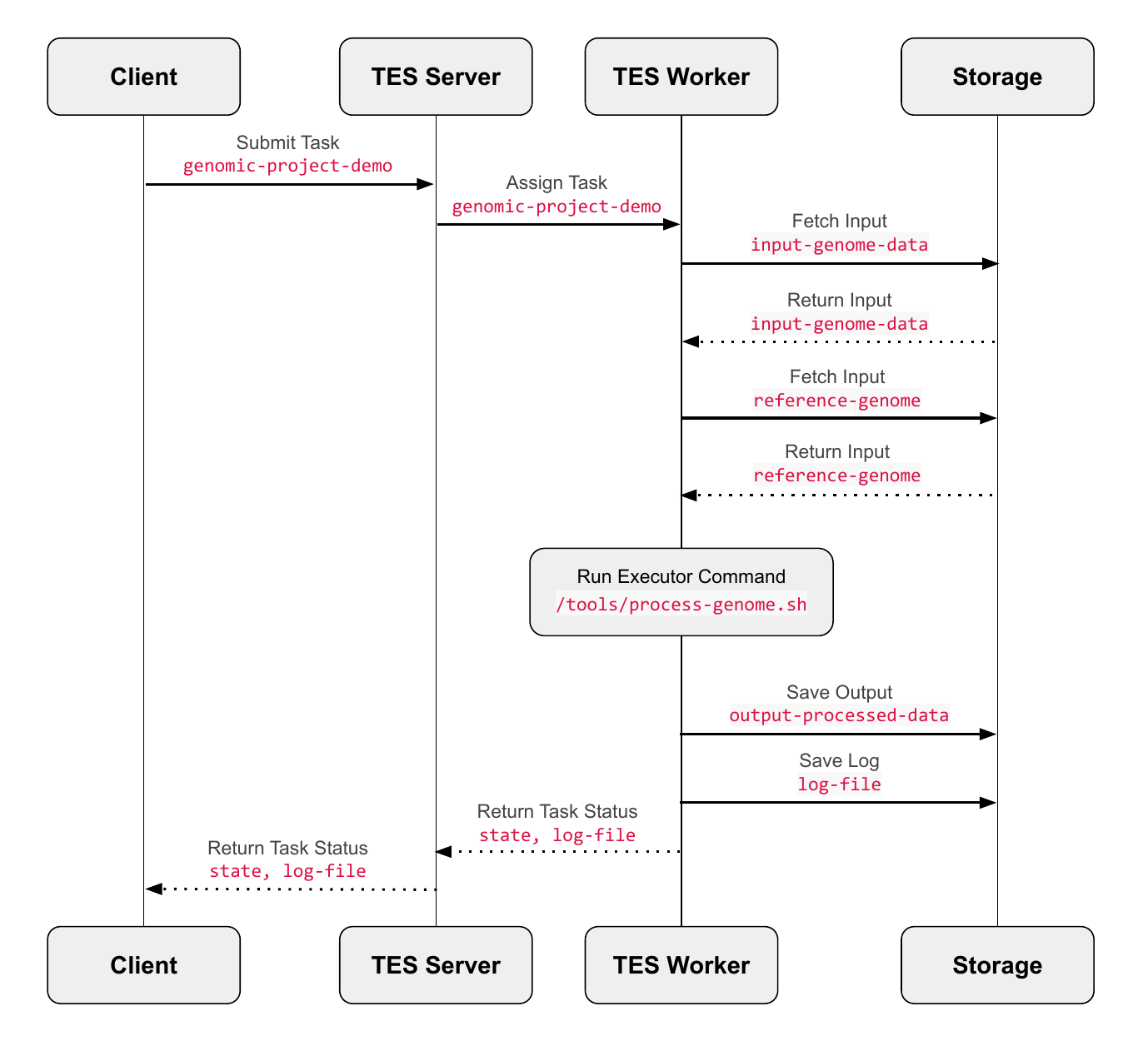}

Figure 3 TES Execution Architecture. An outline of the separate layers found in current TES service implementations. The client talks to a server, which is responsible for allocating a worker node on an HPC/HTC or cloud infrastructure. The TES worker is responsible for transferring inputs, running user code, capturing logging and storing outputs. 
\end{figure}

The TES API is based around the creation, monitoring, controlling and listing of task messages. These have been analogized 
in the TES API as resources under the \verb|/tasks| endpoint, with POST creating new resources and GET returning a paginated 
listing of available resources. Task resources are referenced by a unique identifier which is assigned by the server and returned 
when a resource is created. Using the task identifier, it is also possible to cancel a given task (via POST \verb|/tasks/{id}:cancel|), 
or retrieve detailed information on a task resource (via GET \verb|/tasks/{id}|). Apart from requiring task identifiers to be unique 
for a given TES instance, the TES API does not impose any other restrictions on how these identifiers are formed.

How long tasks are stored is not prescribed by the TES specification and is up to the implementer, or an individual TES instance's 
configuration. When listing available resources, the API provides a mechanism to filter tasks by state or name prefix. When creating 
tasks, clients are allowed to attach an arbitrary set of key/value string pairs to a task that can be used for storage of identifiers 
or any other metadata that clients deem important. These user/client-defined tags may also be used to filter tasks. For the GET methods 
described above, the API allows the user to customize the level of detail for listing into three levels: \verb|MINIMAL|, \verb|BASIC| and \verb|FULL|.

File transfer protocol support in TES relies on external standards. Inputs and outputs are represented as arbitrary strings that an 
implementation then tries to interpret as URLs according to the rules and conventions of supported protocols. To find out which protocols 
are supported by a given TES instance, clients can query the /service-info endpoint via the GET method. The storage property will list 
the supported file transfer protocols. For example, a TES implementation supporting S3 buckets, FTP and HTTP URLs might list s3, ftp and http. 

How TES instances should set up a software environment to execute a command line also relies on arbitrary strings, supplied via 
the image property of a TES executor (see below). It is currently expected that TES instances support URIs pointing to Docker images. 
Explicit support for other containerization frameworks, or even container-independent methods for specifying software 
environments (e.g., Conda) is planned. 

Two key elements of the TES specification, file systems and software containerization, are reliant on external standards. 
The file system can be represented, for example, by a shared file system, an S3 bucket, or FTP HTTP URLs. For containerization, 
the TES API expects a Docker formated container that contains all the software and all dependencies for a given job. In both cases 
the corresponding properties of the TES schema  allow for arbitrary strings to represent elements, and an understanding of the 
capabilities of the TES server implementation. 

A TES task is allowed to have multiple command lines. In the TES API terminology, each command line is defined in 
an executor, and when sending a task request, a client is able to specify multiple executors (specifically, 
the corresponding executors property accepts an array of executors) in an array of \verb|executors|. Each executor 
defines the command line, container image, working directory, stdin/stderr mapping and environmental variables. 
Executors are run sequentially, one at a time. If the \verb|ignore_error| flag is not set (default), a non-zero exit code 
from one of the executors (i.e., command line invocations) will cause the sequence to end early. While this mechanism 
could, in principle, be employed to schedule the execution of linear subworkflows of a given workflow on the same TES 
instance for optimization purposes, this was not the primary intent. Rather, the ability to supply multiple instructions 
for a given task resource  provides an opportunity for clients to drop in setup and finalization methods on the worker 
node, e.g., for increased compatibility with existing workflow management system designs. 

To bridge a crucial gap in the adoption of the TES standard, the TES community recently created a conformance 
test suite for the GA4GH TES API\cite{compliance-tests-ga4gh-tes}. Tests are defined based on a YAML-based specification 
that a dedicated test runner knows how to interpret. This extensible test suite ensures that developers can confidently 
build tools and services that adhere to the API specification, both when developing new implementations and when upgrading 
existing implementations to support new versions of the specification. Currently available conformance tests cover various 
aspects of the TES API, including data structures, messaging protocols, and some functional requirements. The testing 
framework, consisting of the test runner, the test specification and the actual test suite, employs a modular design, 
allowing for targeted testing of specific API components, or the entire API in one go.

\section{Ecosystem}
\label{sec:others}

The goal of TES is to provide an open specification that different groups can utilize, and create a robust ecosystem 
of services that provide and software platforms that utilize it. Ideally, a researcher should be able to easily move from 
HPC to any of the cloud vendors, and not need to change this workflow management system. We measure the success of that goal 
by the number of different implementations that are available and compatible with the API conformance suite (Table 1).

The first element of a robust API ecosystem would be the service providers that implement TES. There are already 
four service providers that officially provide the API to different types of underlying infrastructures.

The Microsoft GA4GH-TES\cite{ga4gh-tes_azure} on Azure project provides a TES API implementation for Microsoft Azure, 
backed by the Azure Batch service. This C\# implementation is one of the core technologies being used to implement the 
Cromwell engine and Broad's Terra platform on Azure. Funnel\cite{funnel} is an implementation of the TES API, designed 
to run tasks on various computing environments such as Slurm, HTCondor, Google Compute Engine, AWS Batch, and others. 
It was developed by a team at Oregon Health and Science University (OHSU). TESK\cite{TESK} is a Kubernetes-native free and 
open source implementation of the TES API, originally developed by EMBL-EBI and now maintained by the ELIXIR Cloud \& AAI GA4GH 
Driver Project, which acts as a batch scheduling system for Kubernetes and compatible Native Cloud solutions like OpenShift. 
Pulsar is a server platform for running jobs from the Galaxy workflow engine on remote systems. They have added support for 
the TES API, so that additional clients that use TES can plug into their network of deployed sites.

Beyond servers that provide the TES API, there is a potential for meta services that are able to provide flexible middleware 
injection to effectively federate atomic, containerized workloads across various environments. For example, proTES is a scalable 
gateway service designed to offer centralized features to a federated network of TES instances, such as serving as a compatibility 
layer for different TES implementations, a head node for smart workload distribution across the network, a public entry point to an 
enclave of private TES nodes, or a means of collecting telemetry. A plugin system allows for the easy creation and injection of middleware 
tailored to specific use cases and requirements, such as access control, request/response processing, or the selection of suitable TES 
endpoints considering legal constraints (e.g., data use) and user/client preferences.

The final element of the ecosystem is the client software that utilizes the API. A number of workflow orchestration 
engines have started to utilize TES as a way to issue required commands as they manage task dependencies. Cromwell is 
an open  workflow engine originally developed by the Broad Institute for the orchestration of bioinformatic workflows. 
It's one of several tools and platforms that are able to run workflows written in the Workflow Description  Language (WDL), 
 Snakemake is a python based workflow management system designed for reproducible and scalable data analysis  that 
 utilizes the TES API to enhance its capabilities. To better support the execution of Snakemake workflows on various 
 compute backends, Snakemake developers have developed a plugin system for such executors for their recent 8.0 release. 
 One such plugin is specifically designed for submitting jobs via the TES API. Nextflow, a groovy based workflow management 
 system designed for portability and scalability, has long been supporting the GA4GH TES API to enable flexible and adaptable 
 task execution across diverse computing environments\cite{Di_Tommaso_Chatzou_Floden_Barja_Palumbo_Notredame_2017}. 
 Limitations towards a more widespread adoption of the TES backend in Nextflow have been addressed on the side of the 
 specification with the release of TES v1.1.0. In the Nextflow workflow management system, developers are currently 
 working towards implementing these changes towards a more powerful TES backend. For the 
 Common Workflow Language (CWL)\cite{Crusoe_Abeln_Iosup_Amstutz_Chilton_Tijanić_Ménager_Soiland-Reyes_Gavrilović_Goble_et_al._2022}, 
 cwl-tes is an experimental workflow engine based on the CWL reference engine that is specifically built around a 
 TES executor. For the Toil\cite{Vivian_Rao_Nothaft_Ketchum_Armstrong_Novak_Pfeil_Narkizian_Deran_Musselman-Brown_et_al._2017} 
 workflow management system, which also supports the execution of CWL workflows, a prototype TES backend has been added recently. 
 Next to workflow management systems, the TES client ecosystem also includes the Python library py-tes for accessing 
 TES servers programmatically (the library is used in some of the workflow management system backends, e.g., the Snakemake one). 
 Finally, a reusable Web Component-based micro-frontend for the TES API is currently being developed as part of the ELIXIR 
 Cloud Components suite\cite{ELIXIR}.

\begin{table}[]
\begin{tabular}{|l|p{2cm}|p{6cm}|l|}
\hline
Type & Project & Description & Source \\
\hline
\hline
Server & Funnel & TES server implementation for HPC/HTS systems & \url{https://github.com/ohsu-comp-bio/funnel} \\
\hline
Server & Pulsar & TES server implementation for the Galaxy/Pulsar federated distributed network & \url{https://pulsar.readthedocs.io} \\
\hline
Server & TES-Azure & TES server implementation for the Microsoft Azure & \url{https://github.com/microsoft/ga4gh-tes} \\
\hline
Server & TESK & TES server implementation for Kubernetes/ Native Cloud systems & \url{https://github.com/elixir-cloud-aai/TESK} \\
\hline
Proxy & proTES & Proxy service for injecting middleware into GA4GH TES requests & \url{https://github.com/elixir-cloud-aai/proTES} \\
\hline
Client & Cromwell & Workflow management system for executing composed in the Workflow Definition Language (WDL) domain-specific language (DSL) & \url{https://cromwell.readthedocs.io} \\
\hline
Client & cwl-tes & Workflow management system for executing workflows in the Common Workflow Language (CWL) DSL & \url{https://github.com/ohsu-comp-bio/cwl-tes} \\
\hline
Client & ELIXIR Cloud Components & Web Component library for interacting with TES services (and other GA4GH APIs) & \url{https://elixir-cloud-components.vercel.app} \\
\hline
Client & Nextflow & Workflow management system for executing workflows composed in the Nextflow DSL & \url{https://nextflow.io} \\
\hline
Client & py-tes & Python client library for interacting with TES services & \url{https://github.com/ohsu-comp-bio/py-tes} \\
\hline
Client & Snakemake & Workflow management system for executing workflows composed in the Snakemake DSL & \url{https://snakemake.github.io} \\
\hline
Client & Toil & Workflow management system for executing workflows composed in the Toil and CWL DSLs & \url{https://toil.readthedocs.io} \\
\hline
\end{tabular}

Table 1: TES ecosystem. A listing of available servers, proxy and client implementations that utilize the TES API.
\end{table}

\section{Discussion}

TES's main goal is to promote interoperability and bring together different computational software 
platforms and infrastructure. Consistent with this goal, the ELIXIR Cloud \& AAI GA4GH Driver 
Project\cite{ELIXIR} is prototyping a cloud infrastructure that uses the TES API at its core to federate 
the execution of computational workflows across different national nodes in the ELIXIR community. Similarly, 
the European Genomic Data Infrastructure, an ambitious project aiming to connect institutions from 20 different 
European countries for the federated analysis of genomics data, from hospitals to research centers, has built 
their federated compute capabilities in their Starter Kit around the TES API. Microsoft initially utilized TES as
 a way to support the Cromwell workflow engine to work on Azure. The implementation was initially rolled into the 
 CromwellOnAzure deployment. With further development, use cases, interest from the community and other workflow engines, 
 the TES implementation was split into its own standalone repository and 
 deployment\cite{ga4gh-tes_azure}.

The philosophy of the API design is to maintain a lightweight interface that allows for multiple implementers 
to easily build compliant clients and servers. The API will continue to evolve, but only in a way that maintains 
this lightweight interface design. In many cases, evolution of the API will not involve changes to the OpenAPI 
description, but rather further refinements to the conformance test suite. One of the main goals of future versions 
is to provide better guidance around authentication, security and software portability. How credentials and authorization 
flows through the various layers of the infrastructure is very important to maintain interoperability of different solutions, 
particularly in multi cloud applications. Passing storage credentials from the user to the node where execution of the job will 
occur needs to be outlined. Additionally, a generalized solution that can pass compute authorizations from users to API endpoints, as 
well as how to restrict compute available to users in hybrid systems, cloud and HPC/HTC, needs to be further developed. 

The TES API was originally designed around docker, with the presumption that the image definition would map to the 
image name passed to a docker command line. However, as implementations and use cases evolved, it became clear that 
other containerization systems, such as LXC, Singularity, rkt, and Podman, would also be seen as possible use cases 
of security and architecture. Additionally, other software management systems such as Conda, Bioconductor and Galaxy 
Tool Shed exist to install and share tools are becoming popular.  From the perspective of the OpenAPI definition, there 
is no alteration required to support these systems as the field would remain a string, the only alteration would be how 
that string would be passed to the deployed system and how the implementer chooses to install and run the software management 
option. How this element of the API affects function will be defined by the conformance tests.

There has been discussion around possible extensions to the API. Ideas such as enabling privacy-preserving federated 
learning, enabling Hadoop or Spark jobs, supporting federated learning use cases and a callback API have all been 
suggested as possible developments for the TES API. These extensions could be added to the API as optional extensions 
that would not be required for basic conformance.

\section{Conclusion}

The TES API is designed to be interoperable, meaning that TES-compatible clients and servers can be developed and used 
independently. This makes it easy to integrate the TES API into existing cloud infrastructures, workflow management systems 
and other relevant software solutions. Importantly, this API allows researchers to develop their analysis workflows independent of 
their infrastructure and deploy them in a multi cloud approach, targeting computational resources that are more affordable or closer 
to the data that needs to be processed. We have already seen a large uptick in the number of groups utilizing the TES API to power their 
systems, both in academia and industry, and demonstrating its utility in powering computational analysis for life sciences. 

\section{Acknowledgements}

This work was funded by grants from the National Institute of Health, including 1U24CA231877 and 1U24HG010263. 
The work was further supported by ELIXIR, the research infrastructure for life-science data. Additionally, 
this work was supported by Microsoft's Biomedical Platforms and Genomics Team. Working group leadership includes Ania 
Niewielska (until 2022), KE, AK, and VM. Additional API contributors include Susheel Varma, Adam Struck, Alexander Buchanan, 
Angel Pizarro, Jeena Lee, Ryan Spangler, Jeff Gentry, Brian O'Connor, and Benedict Paten.

VM and KE are co-champions of the GA4GH TES API specification. AK and MHM are frequent contributors. 
Each of the authors have contributed significantly (leading role in planning, implementation and/or maintenance) 
to one or more of the TES ecosystem projects described in this manuscript. AK: TESK, proTES, TES compliance test 
suite, ELIXIR Cloud Components. KE: Funnel, TES compliance test suite, cwl-tes and py-tes. MHM and VM: Tes-on-Azure.

Finally, the authors would like to thank the GA4GH leadership, secretariat and steering committee for adopting 
TES as an official GA4GH standard and help with project administration and maintenance

\bibliographystyle{unsrtnat}
\bibliography{references}

\end{document}